\documentclass[aps, twocolumn,showpacs,preprintnumbers,amsmath,amssymb,superscriptaddress]{revtex4}
\usepackage{graphicx}

\usepackage{color}

\begin{document}

\title{Simulation of Droplet Trains in Microfluidic Networks}

\author{Mehran Djalali Behzad}
\affiliation{Department of Physics, Sharif University of Technology (SUT),  P.O. Box 11365-9161, Tehran, Iran}
\author{Hamed Seyed-allaei}
\affiliation{Department of Physics, Shahid Beheshti University, G.C., Tehran, Iran}
\altaffiliation{Institute for Research in Fundamental Sciences (IPM), P.O. Box 19395-5531, Tehran, Iran}
\author{Mohammad Reza Ejtehadi}
\affiliation{Department of Physics, Sharif University of Technology (SUT),  P.O. Box 11365-9161, Tehran, Iran}

\begin{abstract}

In this work we show that in a microfluidic
network and in low Reynolds numbers  a
system can be irreversible because of hysteresis effects.
The network, which is employed in our simulations,  is taken from recent experiments. The network consists of one loop connected to input and output pipes. A train of droplets enter the system at a uniform
rate, but they may leave it in different patterns, e.g. periodic or
even chaotic. The out put pattern depends on the time
interval among the incoming droplets as well as the network geometry and for some parameters the system is not reversible.

\end{abstract}

\maketitle

\section{Introduction}


Fluids behave differently at the microscale.
In this scale, factors such as surface tension, energy dissipation, and fluidic resistance are dominant and the flow stays laminar because of the small Reynolds number.
These factors lead to unintuitive behavior in a microfluidic network which is a network of channels and pipes about one to hundreds of micrometer large~\cite{squires2005mfp, whitesides2006oaf}.
Microfluidic devices have applications in inkjet printers, lab-on-a-chip devices, fuel-cells~\cite{choban2005mlf}, biochips~\cite{schena1995qmg,ko2003pbm} and recently, the possibility of making logical gates and signal encoders and decoders has been studied~\cite{Bubble.Logic.Scince, Coding_Decoding.Science}.

\begin{figure}
\centering
\includegraphics[width=6cm]{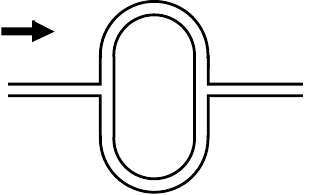}
\caption{\label{fig:setup} The configuration of pipes in the experiment~\cite{Coding_Decoding.Science} and our simulations. From point $A$ to $C$ there are two ways, $ABC$ and $ADC$. When the lenght of these two pathes are equal we call it a symmetric setup and when they are different, we name it asymmetric setup.}
\end{figure}

In a recent work Fuerstman {\it et al.} look at the possibility of coding and decoding of signals using a simple network of microtubes which is made of just one single loop with two connecting pipes~\cite{Coding_Decoding.Science} (FIG. \ref{fig:setup}).
A constant flow of fluid enters the system from one side of the device (input pipe) and leaves from the other end (output pipe).
This fluid is used as a medium to carry bubbles or droplets of an immiscible liquid.
The droplets are large enough to block the pipes but small enough to be stable along the path.
Reaching a T junction the bubbles must go only through one of the ways since they can not split.
The decision depends on the pipes flux velocities which are history dependent and are related to the distribution of other droplets which have already arrived into the system.
A droplet has the same cross section as the pipe, with a very small length, and in each cross section of the pipes, we have exclusively a droplet or the carrier fluid.
Therefore, we can see droplets as binary signals which carry information on a suitable medium, and microfluidic network can be represented by equvalent electrical circuits~\cite{ajdari2004sfn}.

The fluid input flow in the system is in a constant rate, but the flow rate in any branch is a function of the loop geometry (length and  diameter) and  more interestingly, the presence of droplets in either branches.
While the capillary force of a droplet slows down the current in one branch, the other branch may become more favorable for the next coming droplet.
Considering the input stream as a binary signal (droplet means 1, no droplets means 0) droplets get different patterns(signals) in the output with respect to the time intervals between entering and the geometry of the loop. This can be seen as a way to code a signal.
The reversible dynamics of the fluids at low Reynolds numbers suggests that the signal could be decoded if the out put signal is  fed to a system of pipes with the same geometry~\cite{Coding_Decoding.Science}.
The possibility of making logical microfluidic devices has also been examined by Prakash and Gershenfeldof based on the same concept~\cite{Bubble.Logic.Scince}.
They have employed more complex networks to demonstrate how such a  systems may act as logical gates.
The possibility of microfluidic logic has raised also the interest of computer scientists~\cite{1284650}.

Tracking the droplets inside the system in addition to time intervals between the output droplets, provides us information about the branch that a droplet selects.
This gives us another informative binary signal which has been studied by Jousse {\it et al.}~\cite{Bifurcation.of.droplet.PRE}.

We run a simulation based on the experiment~\cite{Coding_Decoding.Science} to investigate the relation between the time intervals of the input signal and the pattern of the output. 
The simulation results however show that we are in low Reynolds number regime which may show chaotic behavior because of the nonlinear history dependent dynamics of the system. Besides, the reversibility is also not a general property of the system and it may break even in non chaotic regime. We also see how related are the periodic patterns of time intervals and the path selection signals.

\section{The model}
\label{Model}

Our microfluidic model network is simulating the experimental network which has been introduced in reference~\cite{Coding_Decoding.Science}.
The flow enters the system through the input pipe.
It reaches the first T junction and splits into two branches which join again at the second T junction.
The two branches of the loop are identical except in their lengths.
In fact, the length difference of the branches of the loop is one of our model free parameters(figure \ref{fig:setup}).

In each step of the simulation, we update the flow speed in all the pipes. As the geometry does not change during the simulation, the only relevant parameter which controls the flow velocity is the number of the droplets in the pipes.
We can then simply follow the droplets in the pipes until the next upcoming event, in which one of the droplets reaches one of the  bifurcation points.
We recalculate the flux rates of the new configuration and continue in this way whether the droplet leaves or enter the loop. 
 In a cylindrical pipe of length $L$ and diameter $D$, at low Reynolds numbers, the flux $Q$ is related to pressure difference between the pipe ends, $\Delta P$, via
\begin{equation}
\frac{\Delta P}{Q} \sim \frac{\mu L}{D^4},
\end{equation}
where $\mu$ is the liquid viscosity.
In the model, we also need to consider the effect of the droplets on the current. A droplet affects the flux velocity by introducing a capillary pressure drop,
\begin{equation}
\Delta P_{\rm Ca} \sim \frac{\sigma}{R}\, \, Ca^\frac{2}{3},
\end{equation}
where $\sigma$  and
\begin{equation}
Ca = \frac{\mu v}{\sigma}
\end{equation}
are surface tension between  the gas and the liquid and capillary number, respectively~\cite{Equation1, Equation2, Equation3}. Here $v$ indicates the droplet velocity which is the same as that of the carrier fluid.

According to the above equations, we are able to calculate the pressure difference in any branch.
Flux conservation on the network nodes results in a system of cubic equations for the flux rates which could be solved numerically.
The flux rates of the pipes are preserved as long as the number of droplets in all the pipes are unchanged.
Therefore, we calculate the position of the droplets over the time until a new droplet enters a pipe or a moving droplets meet the end of a pipe (we call it a new event).
When a droplet reaches the bifurcation point, it selects the branch that has a stronger flow and for the case of equal flux (in the range of our numerical accuracy),
 we suppose that one of the branches is always more favorable.
Then we recalculate the pressure differences and the flux rate  after any event and continue in this way.

In our simplified model, we study the effect of the length difference $\Delta L$ between the branches on the output signal.
We also investigated the impact of time interval of input droplets, $\Delta T$ on the output signal, while all other system parameters are fixed.

We perform the simulation for two different pipe settings simulating the original experiment.
In one of them branches  have the same length ($\Delta L = 0$) with a ratio $L/D = 40$  (symmetric).
In other one there is a difference of $10\%$ between the lengths with the same diameters (asymmetric).
For both symmetric and asymmetric configuration we study the time coding of the output signal as a function of the time intervals between entering droplets.

To study the reversibility of the process we consider two different ways of signal decoding.
In the both methods the sequence of the time intervals between the output droplets is saved and used as input signal in the reversed process.
In one of the methods which we call in-place decoding, we stop the coding process and reverse the direction of the movements.
For instance, if the input flux is entering from the left and exiting from the right,
we reverse the flux direction from right to left and the distribution of the droplets inside the system also is preserved at the time of the flux reversion.
In the other decoding method we suppose that we have only the coded signal  and a decoding device, identical to coding device.
That means we feed the coded signal to an initially empty system with exactly the same geometry of the coding device but in the reverse order.
From the application viewpoint, this method may be more interesting and in contrast with the first method we call it out-of-place decoding.


\section{The results}
\label{Results}

In the simulation we enter the droplets in an uniform flat rate.
Then an input signal is simply specified by a single parameter $\Delta T_i$, the time interval between entering droplets.
But in the output the time intervals between the droplets might be different.
Then the output signal is specified by a sequence of $\Delta T_n = T_n - T_{n-1}$ where $n$ counts the  droplets.

\begin{figure}
\centering
\includegraphics[width=4cm]{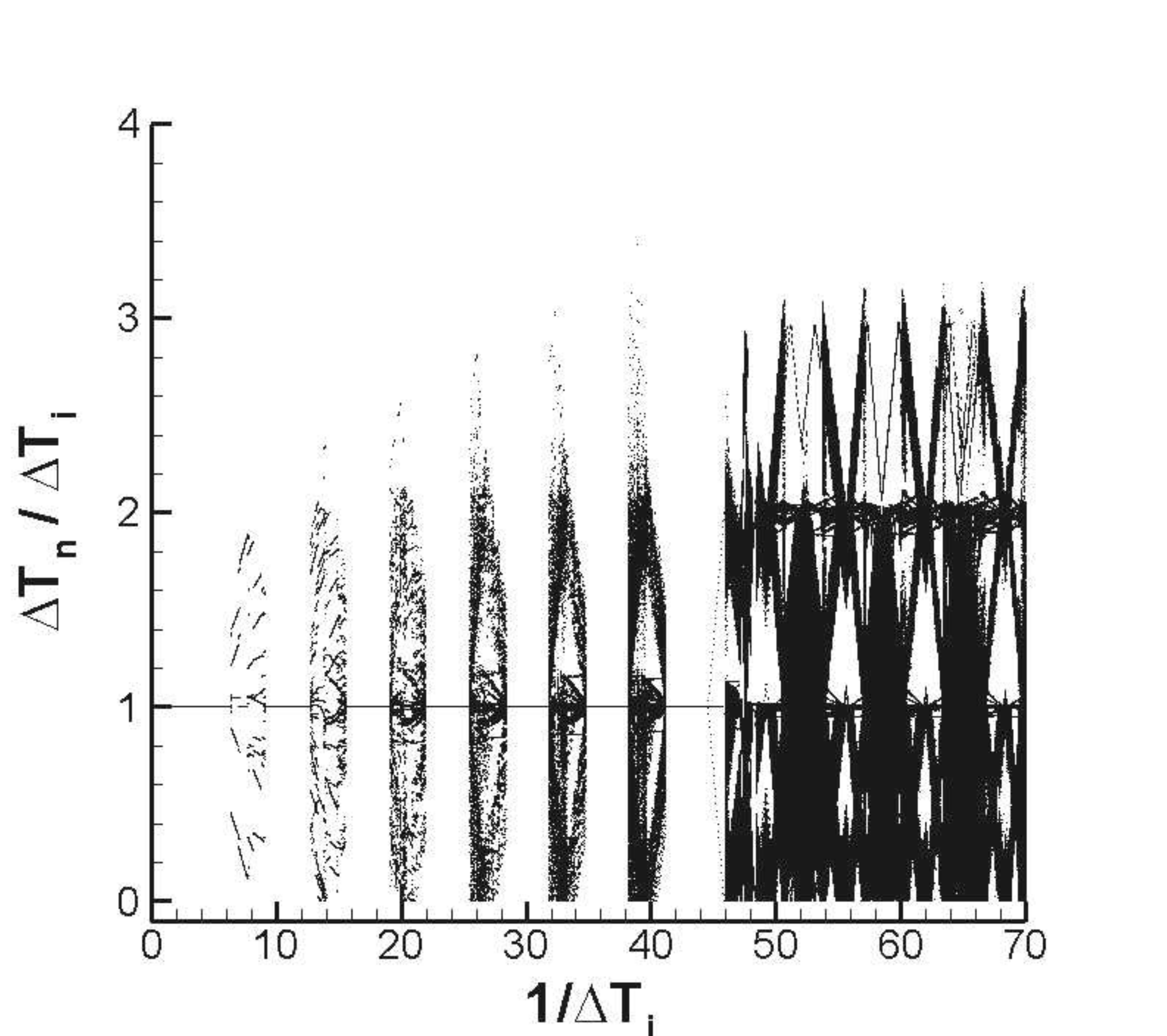} 
\includegraphics[width=4cm]{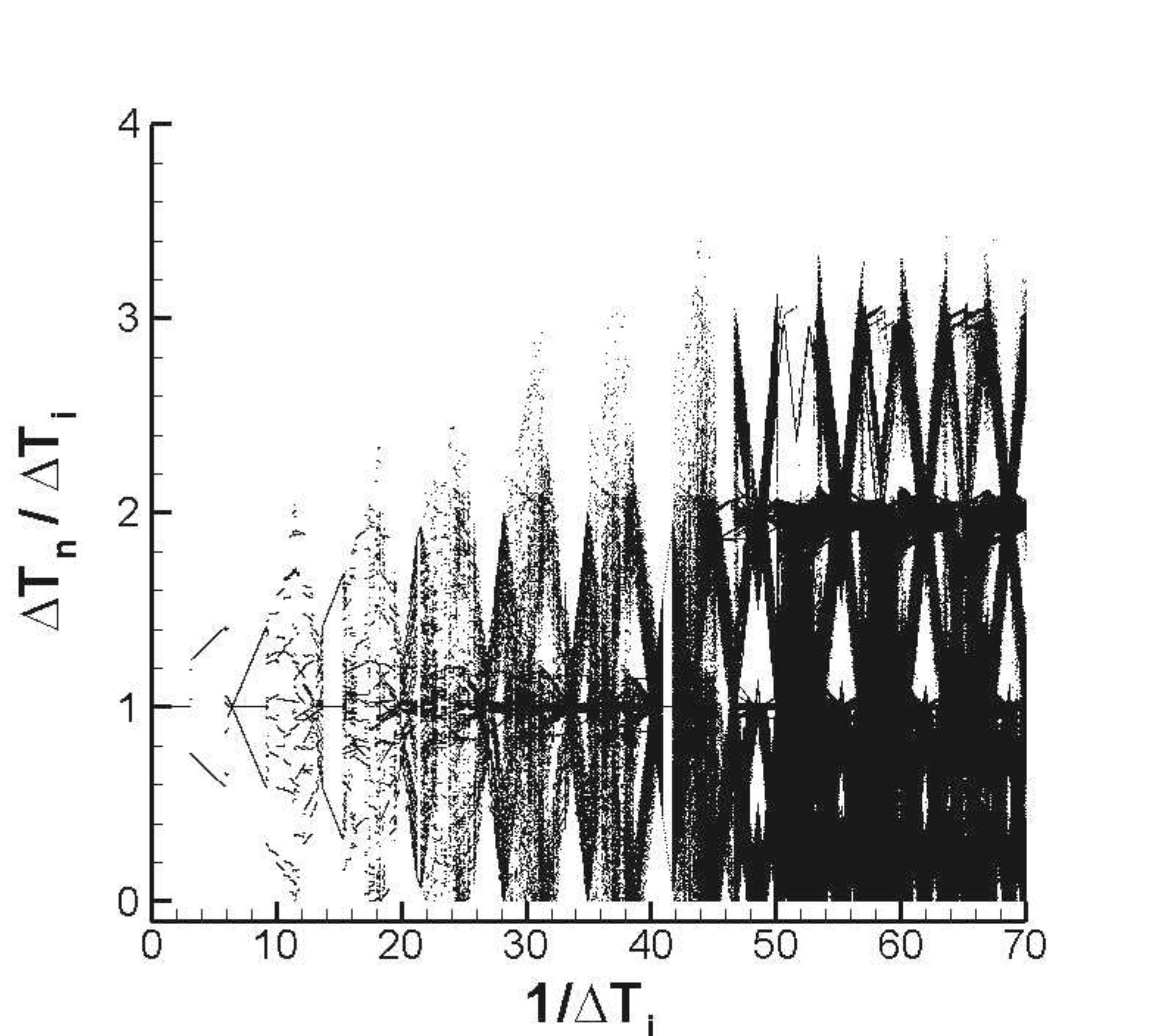}  
\\
\includegraphics[width=4cm]{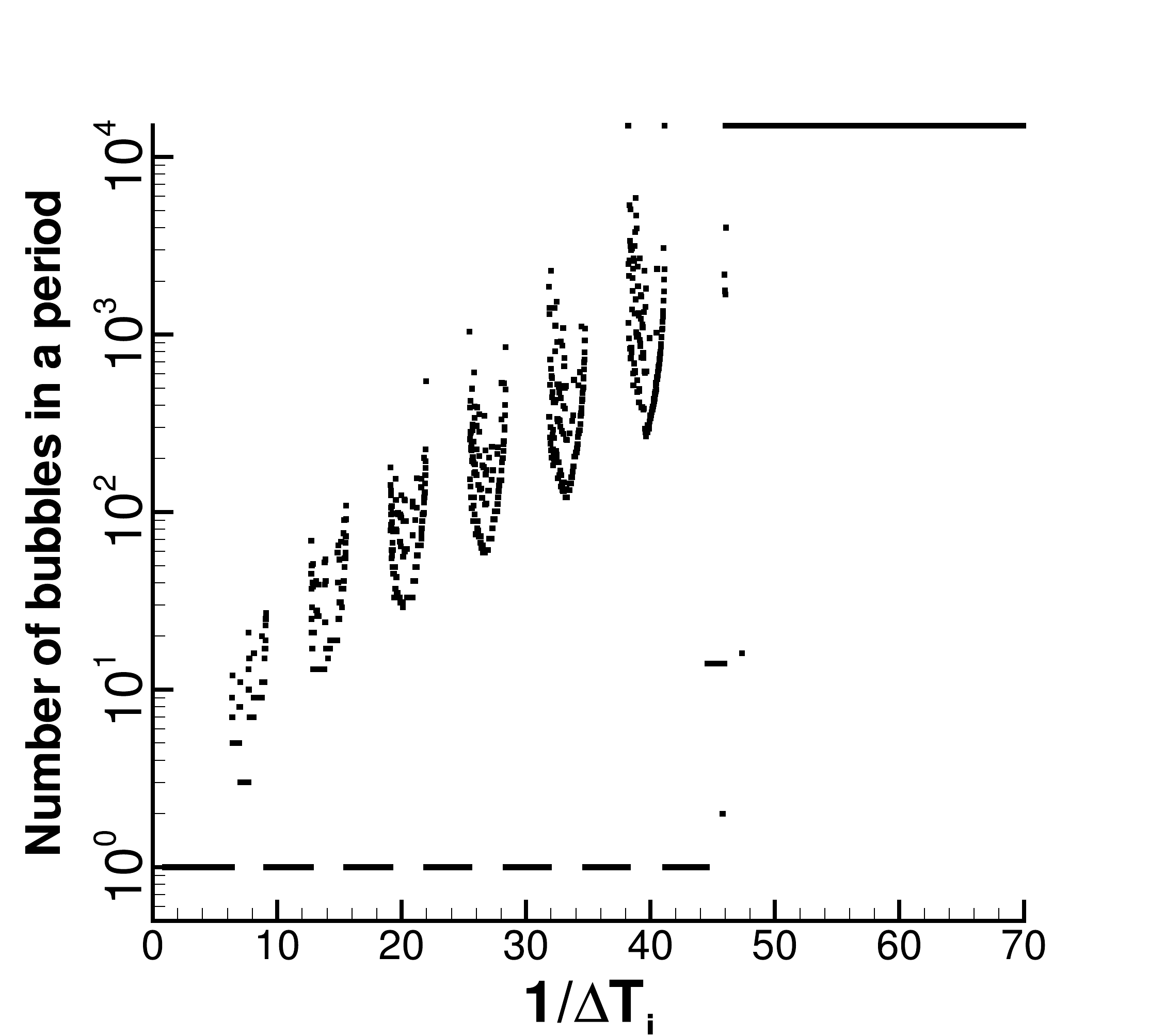}
\includegraphics[width=4cm]{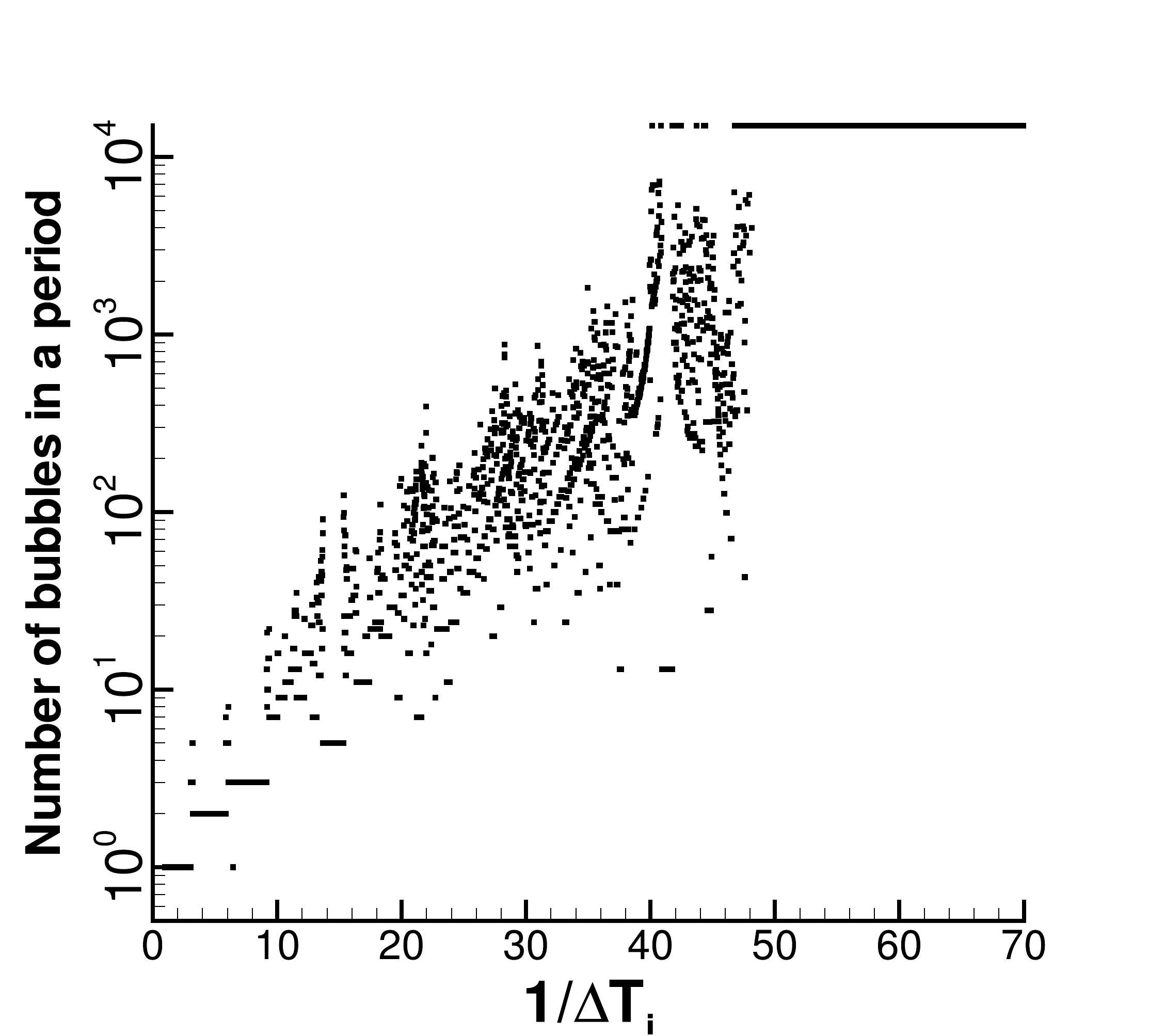}\\
\caption{\label{fig:t_t} \label{fig:periods}
Output time intervals versus input time interval (top) for symmetric setup  (left), with two identical branches, and asymmetric setup (right), in which,  one branch is longer.
The period of the output signal as a function of the input time interval is shown below each case.}
\end{figure}

For large enough $\Delta T_i$'s it is trivial that the output pattern is identical to the input pattern.
By decreasing $\Delta T_i$, at a certain point, we start to get different patterns at the output.
This happens if a droplet reaches the entering T junction, while the former droplet has not left the loop yet.
So the droplet at the junction will select the other branch which is empty and has a higher flow rate.
At this point we start to get a periodic output pattern (periodic sequence of $\Delta T_n$s) that are different from the uniform input signal.
The period length (the number of droplets in a periods) increases as we decrease $\Delta T_i$.
By even more decreasing $\Delta T_i$ for some smaller time intervals, we start to get disordered patterns with no visible order or periodicity on the output (FIG.\ref{fig:t_t}),
there, the period length reaches the number of droplets involved in the system which means the system is either aperiodic (chaotic) or has a period longer than the simulation time.

Similar to many other chaotic systems,
here we also observe that the system reemerge into the chaotic regime in different points for the both symmetric and asymmetric models.
Interestingly, there are some intermediate intervals where, the  coder loop is invisible and the output and input are identical.
The period of the output signal as a function of input time interval is shown in FIG.\ref{fig:periods} and FIG.\ref{fig:t_t}.


As it is shown in figure~(\ref{fig:t_t}), 
similar behavior has been observed in both symmetric and asymmetric networks however the figures are different. 
This is indicating that even at low Reynolds numbers, the system shows chaotic behavior merely as a result of history dependent dynamics of the model. 
The same feature have been observed in history dependent ecological  models~\cite{gerami2000hds}.

Similar to many other nonlinear systems, the output signal may change in time and we need to wait  until it converge to its attractor. 
To sample the output intervals, we wait until a stationary distribution function of the intervals is achieved.
This might take long time.




We study the reversibility (decoding) of the process in both in-place and out-of-place methods.
It is not surprising that in the chaotic regimes, the process is decodable with neither methods.
It shows that in the chaotic regime the input signal information is lost and it can not be decoded.
Instead it is possible to restore the original signal when we have periodic outputs however it is not always true that all the processes are reversible when they are not chaotic. This is more surprising for the case of in-place decoding.

\begin{figure}
{\large a \hfil b} \\
\includegraphics[width=4cm]{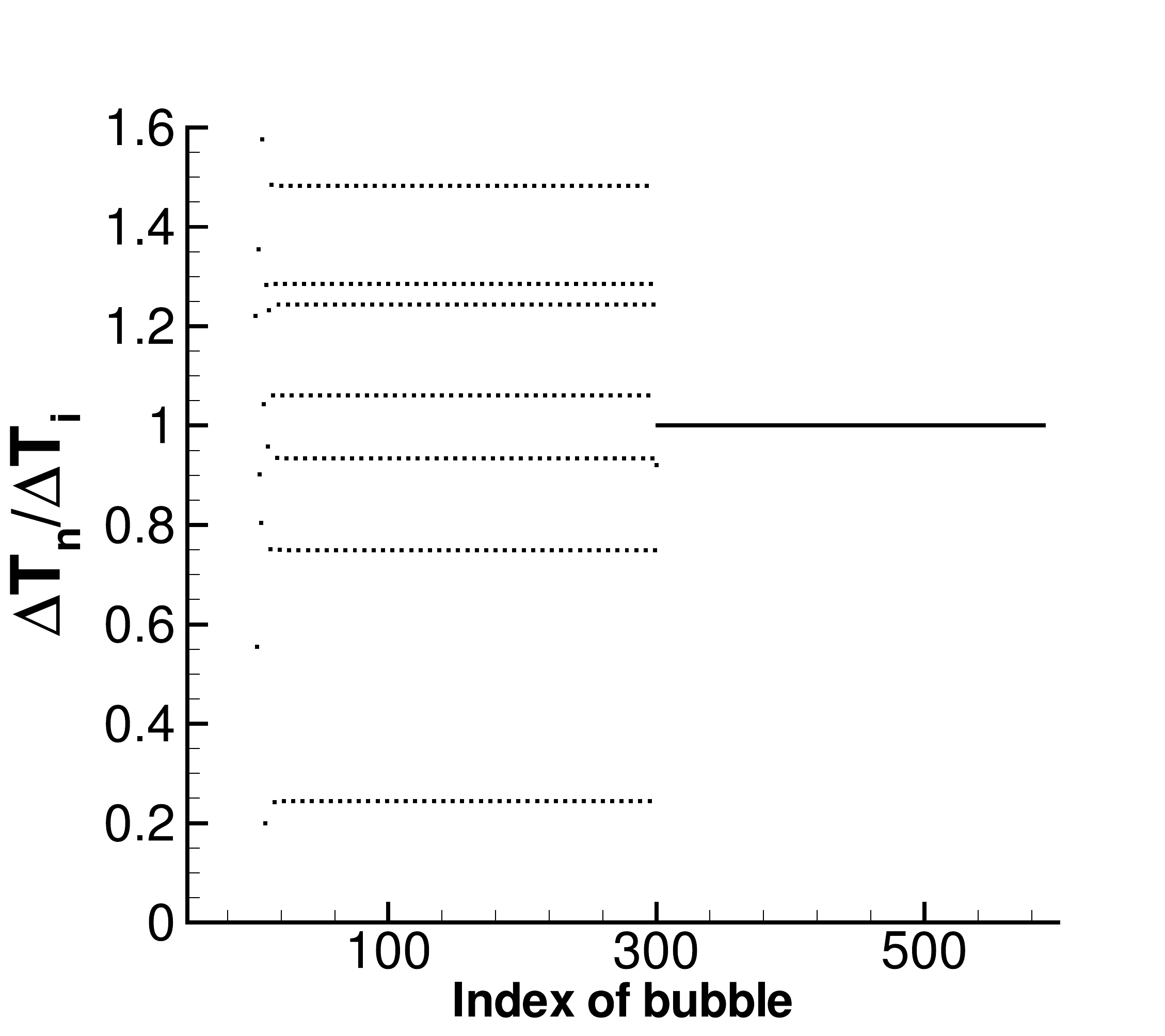}
\includegraphics[width=4cm]{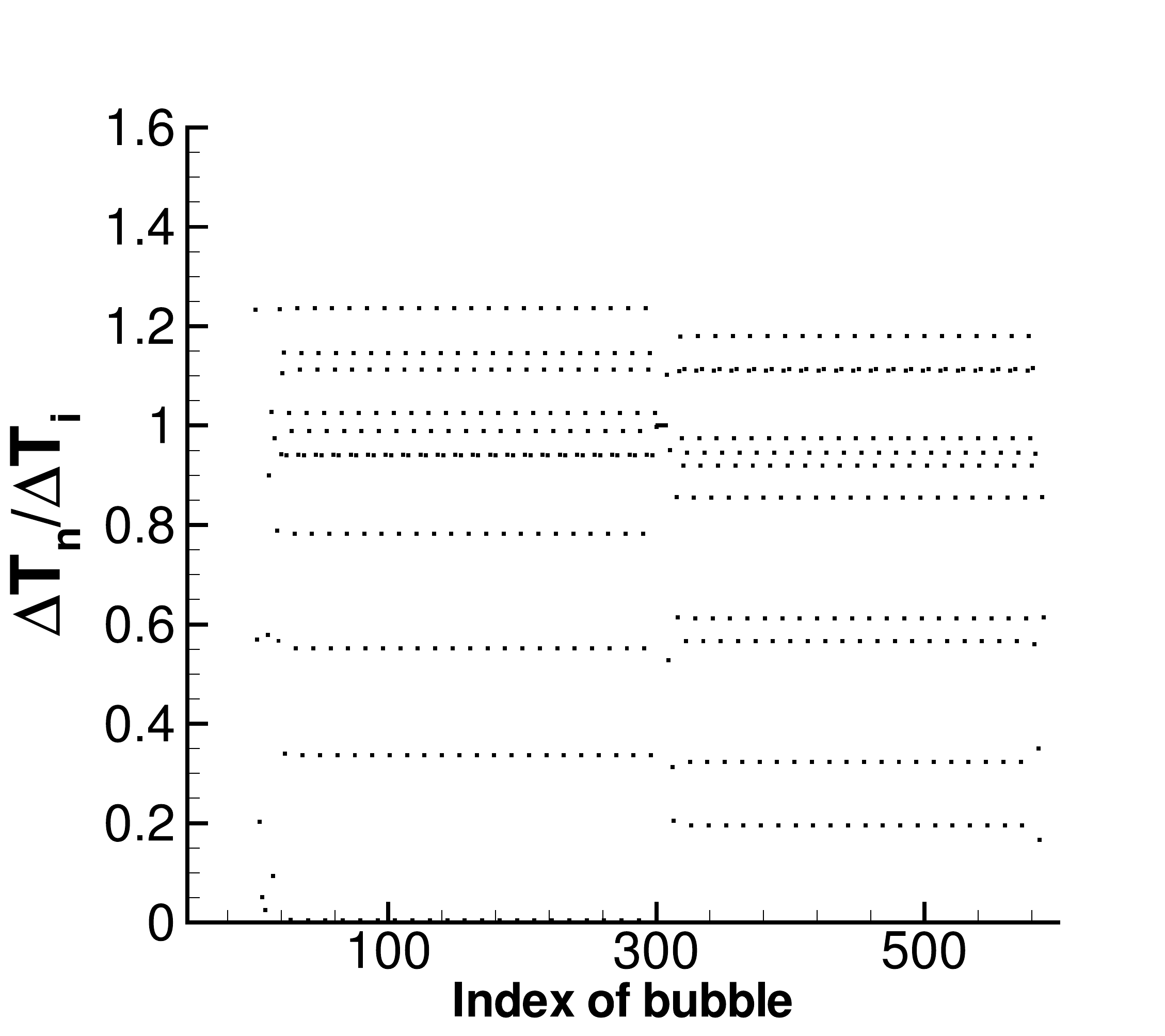} \\
\vspace{0.5cm}
{\large c \hfil d} \\
\includegraphics[width=4cm]{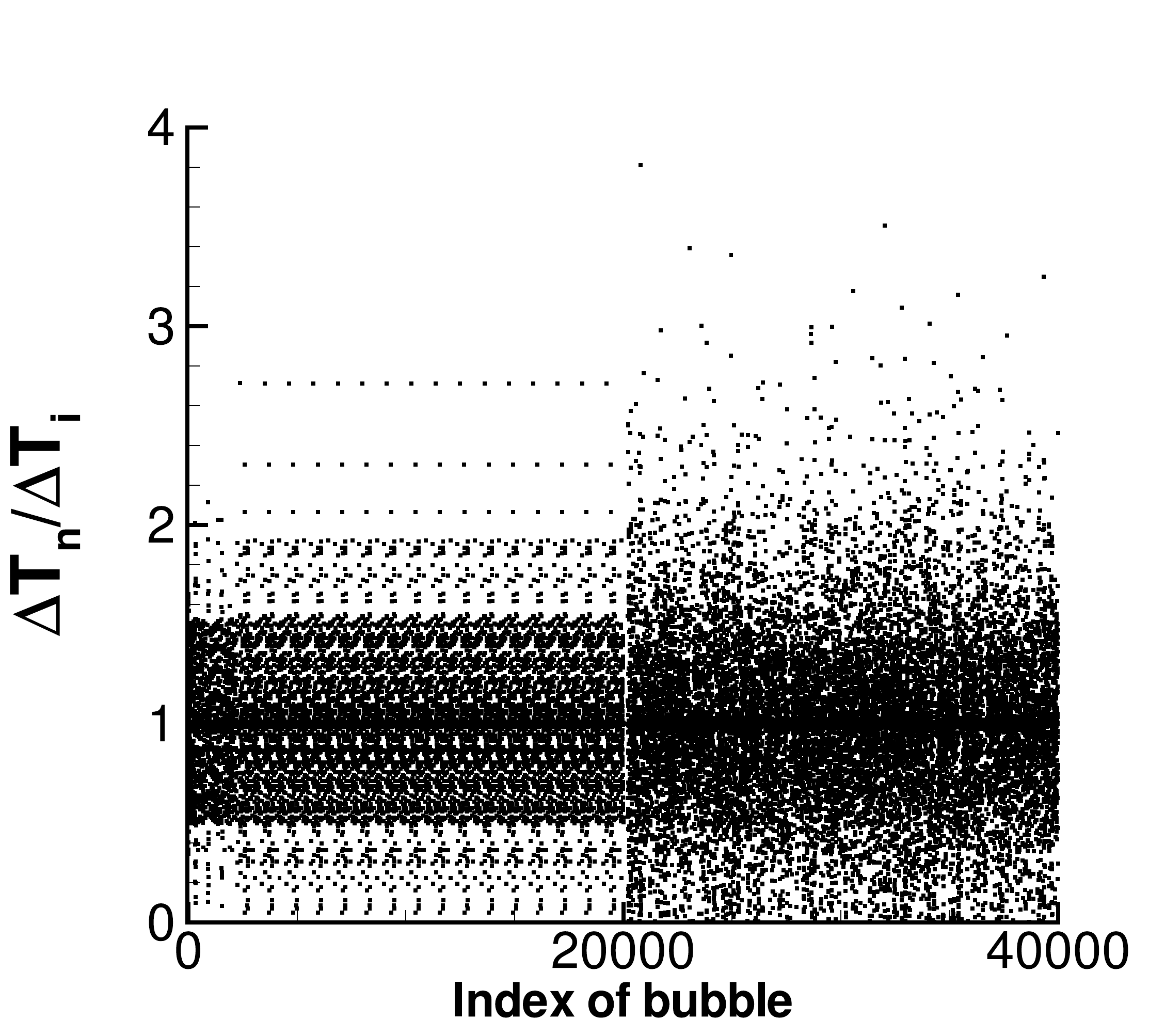}
\includegraphics[width=4cm]{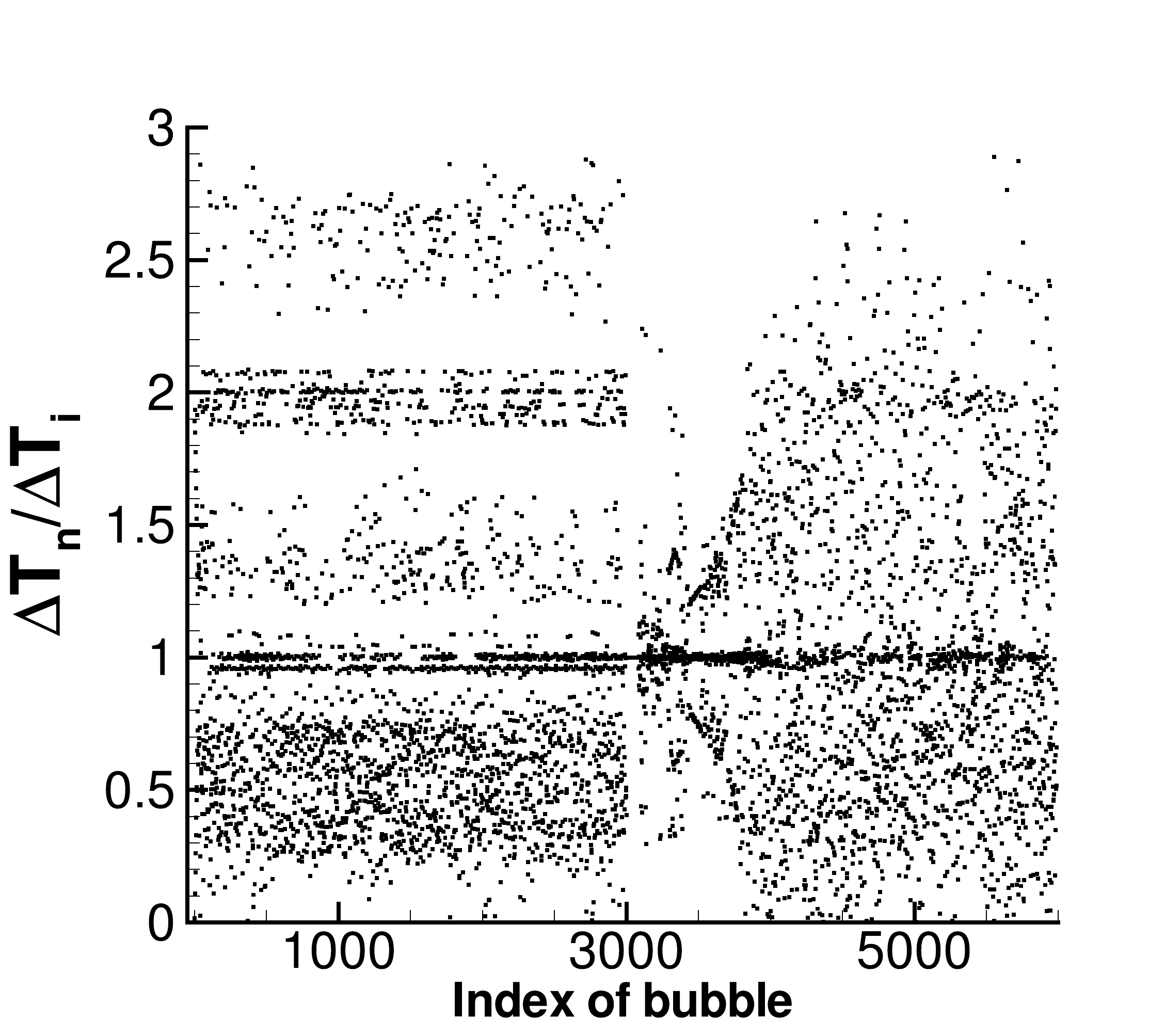} \\
\vspace{0.5cm}
{\large e} \\
\includegraphics[width=6cm]{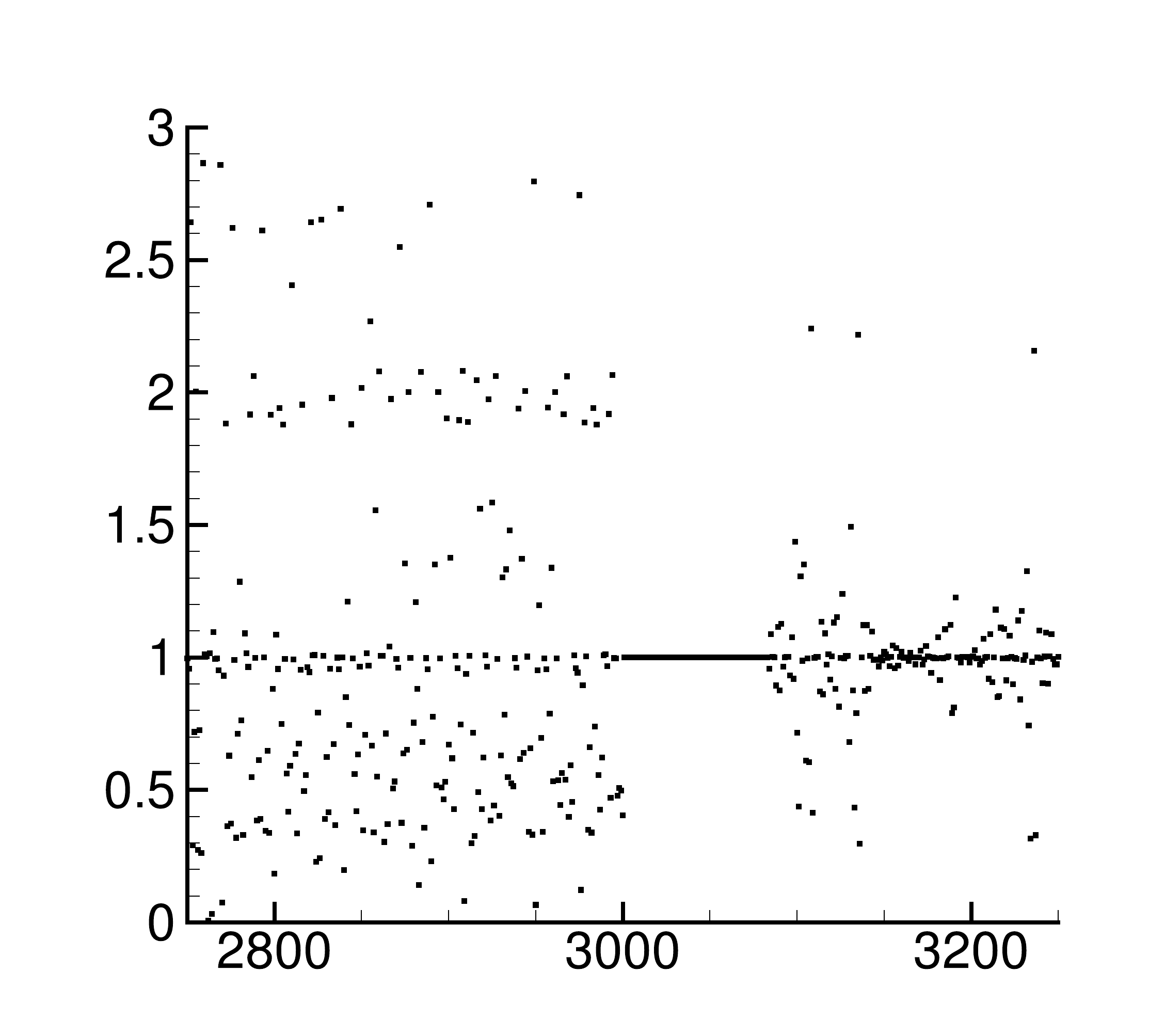}

\caption{ \label{fig:index_time} In-place decoding for asymmetric model: the time intervals are shown for ``before'' and ``after''  the reversion of the flow for 300 (a and b),20000 (c) and 3000 (d) droplets,
in the left and the right side of each graph (see text).
For $\Delta T_i=1/10.00$ (a), there is a regular and periodic coded pattern which is decoded perfectly.
For $\Delta T_i=1/11.45$ (b), there is  again periodic coded output, but the time reversal pattern is a different periodic signal.
For $\Delta T_i=1/42.60 $ (c),
however, the coded signal is periodic, in reversion there is no attractive pattern and the system shows chaotic behavior very soon.
For $\Delta T_i=1/54.00 $ (d and e), the output pattern is aperiodic with the values concentrated on some strips.
The reversion produces a similar pattern with more fluctuations.
}
\end{figure}

The different behaviors of the system for some input time intervals are presented in figure~\ref{fig:index_time} for asymmetric setup.
In the figure the patterns of time intervals are shown for ``before'' and ``after''  the reversion of the flow.
First, the flow goes normally through the network and the output is shown for some droplets (the left part of each graph),
then we look at the time interval between these droplets when the flow is reversed (the right side of each graph).
The graphs differ only in the time intervals of the input signals.
Therefore, as it is shown in the figure, even for periodic output signals, the initial signal sometimes is reconstructable and some times not.
It is also possible that a periodic encoded signal shows chaotic behavior in its return (figure~\ref{fig:index_time}c).
The similar behavior is observed for symmetric setup also.
Here as we are at low Reynolds numbers, time reversality might be expected.
But again history dependence of the process plays its role. 
A droplet chooses a branch that has a faster flow. But when it leaves the branch, the flows might have been changed. As a result, when we revers the time, the droplet who has just left a branch, might decide to go in the other branch for it has a faster flow.


\begin{figure}
\includegraphics[width=4cm]{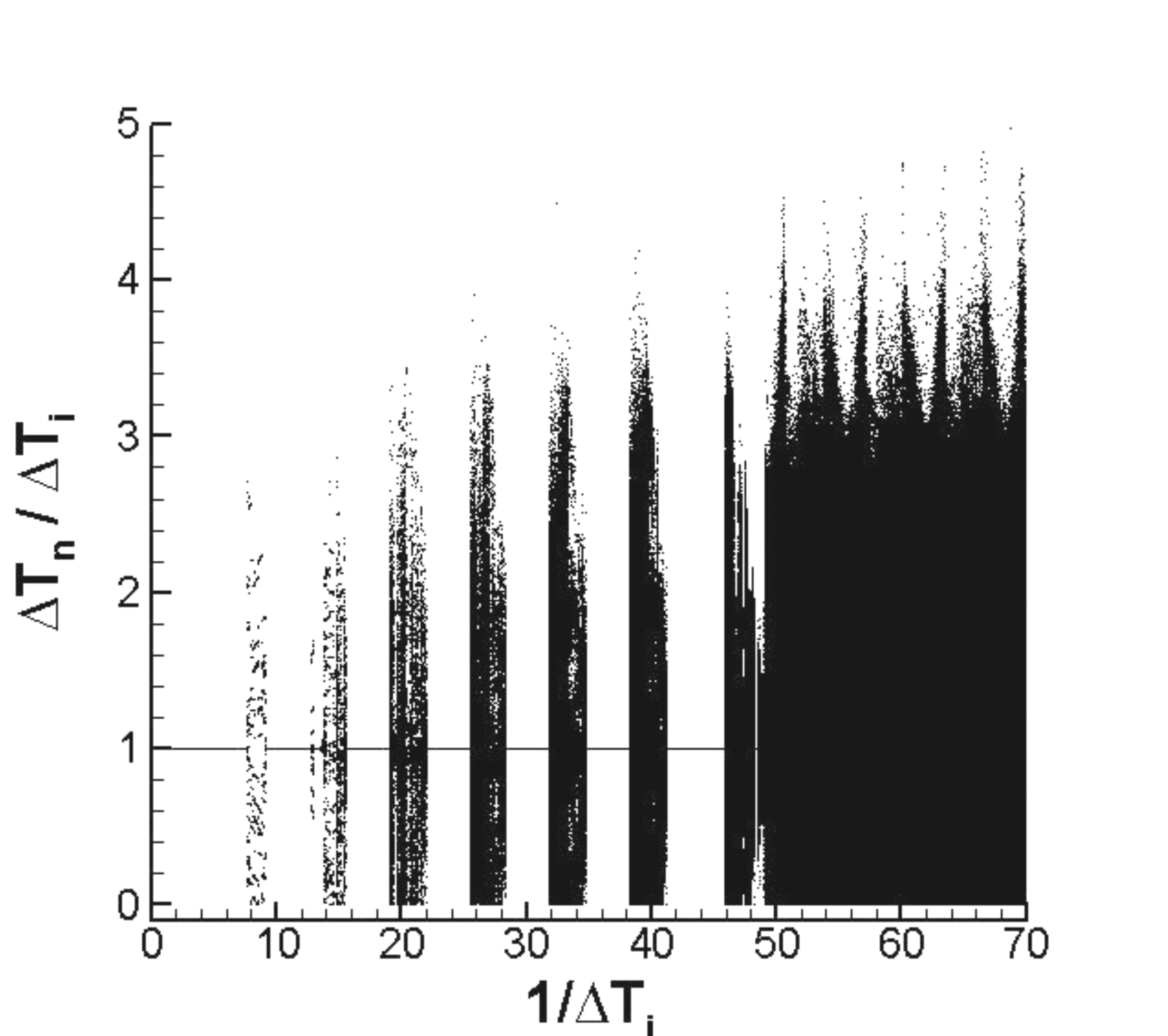} 
\includegraphics[width=4cm]{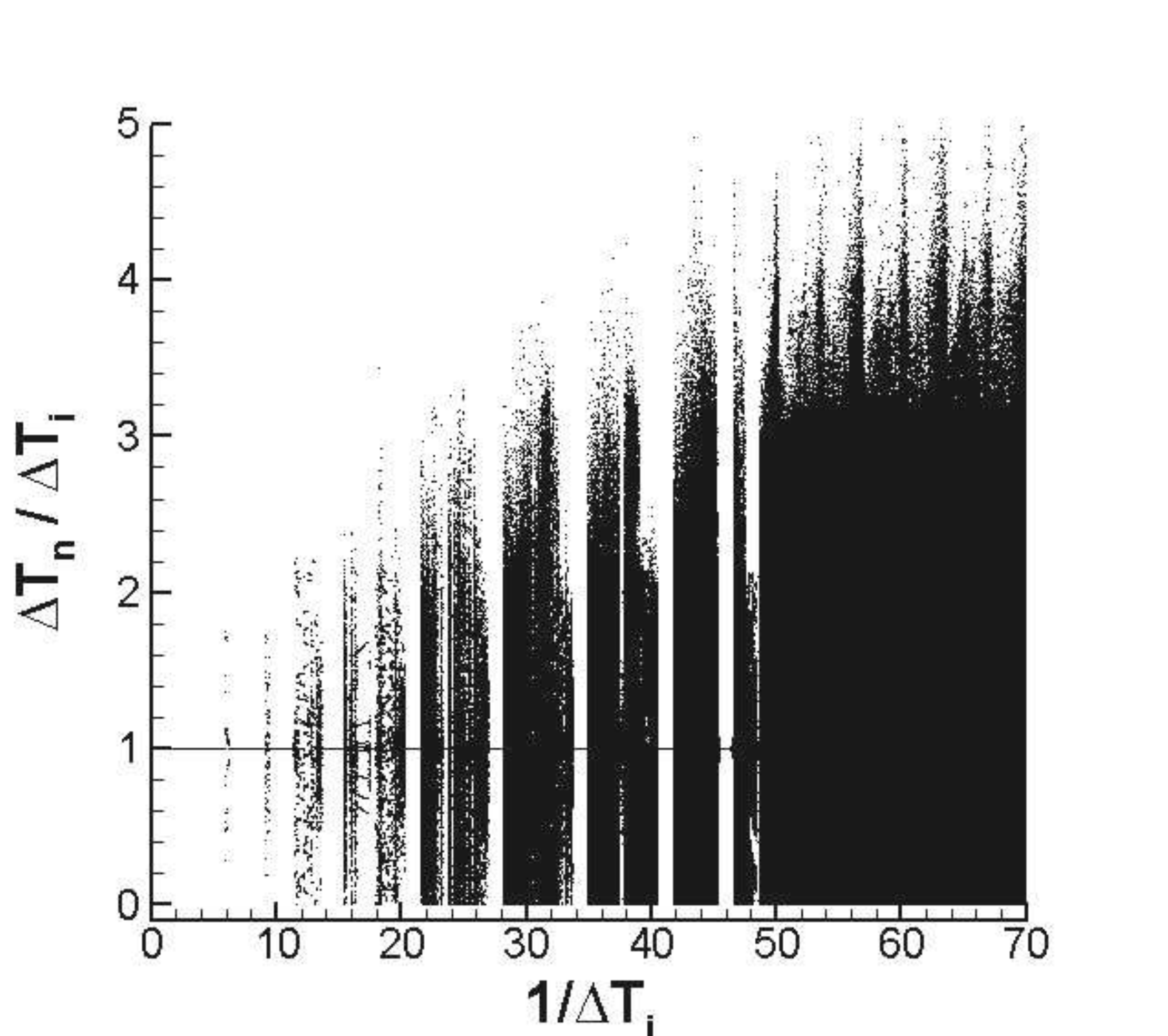} 
\\
\includegraphics[width=4cm]{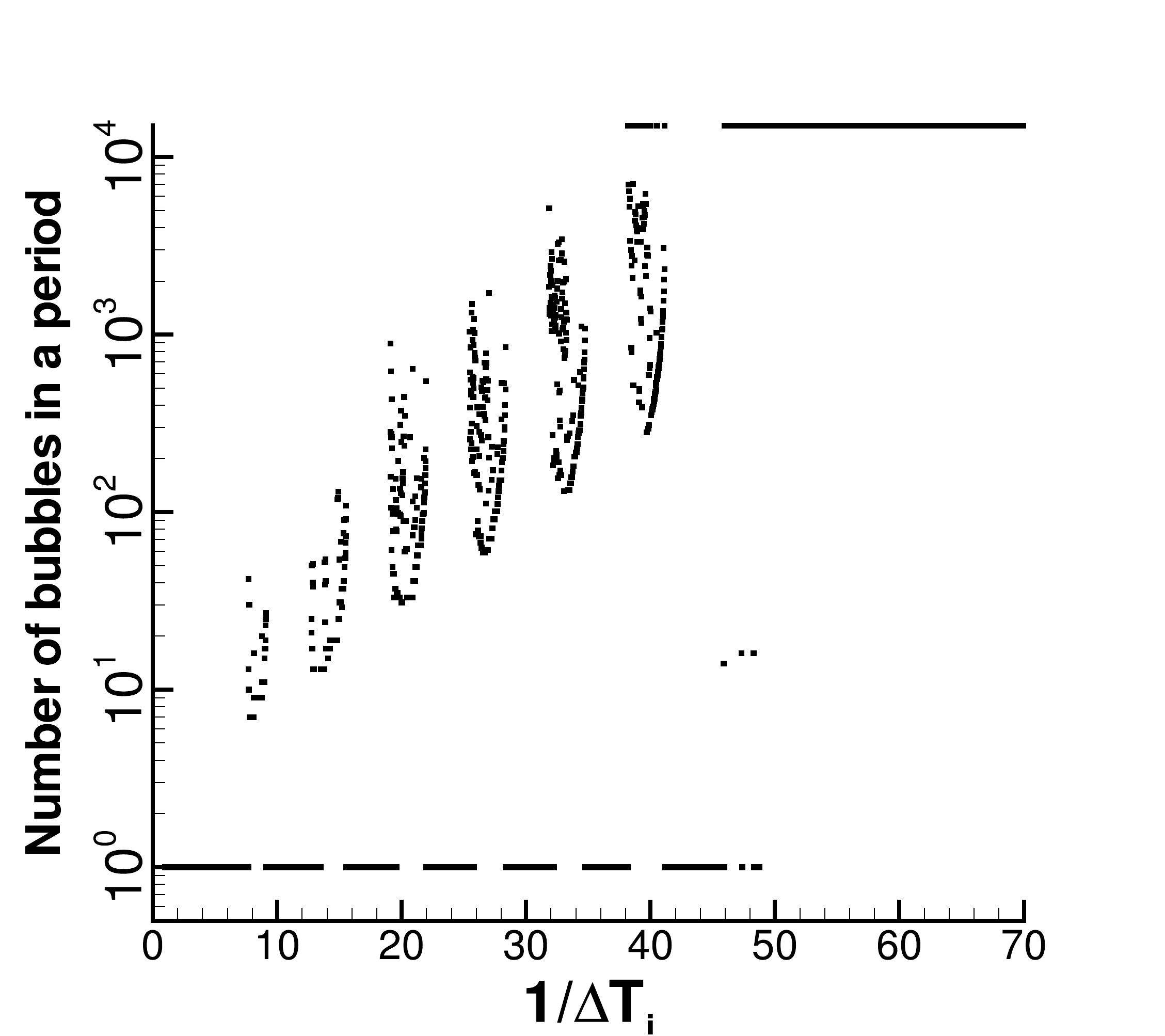}
\includegraphics[width=4cm]{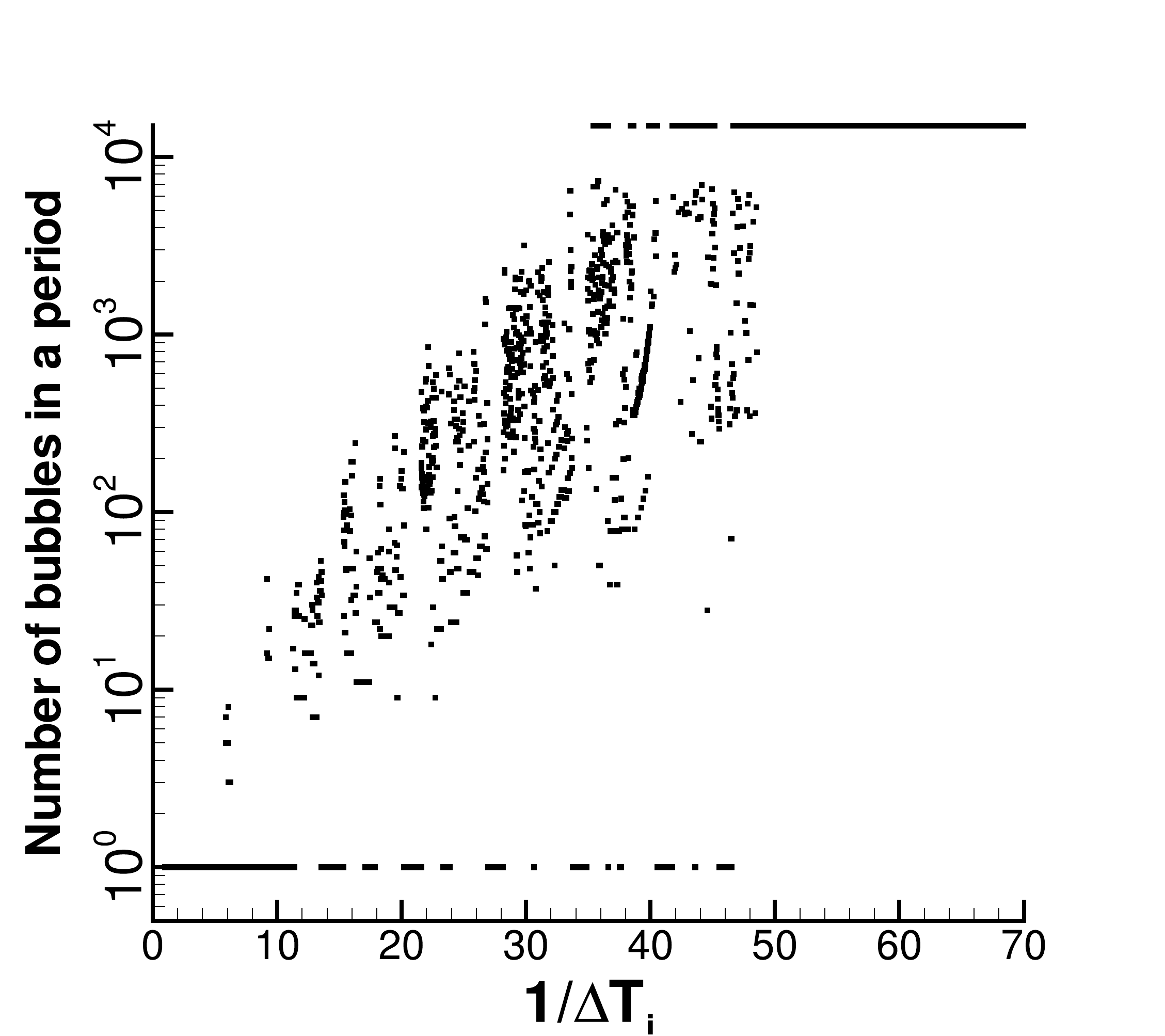}

\caption{\label{fig:decode}
Time intervals between the droplets of in-placed decoded signals as a function of uniform time intervals between the droplets in original signal~(top) and the period of the output signal as a function of the input time interval~(bottom). .
For both symmetric~(left) and asymmetric~(right) setups there are some regions where the signal is decoded successfully.
The corresponding encoded signals are shown in figure~\ref{fig:t_t}.}
\end{figure}

Looking at figure~\ref{fig:index_time} one might ask if the decoded signals are stationary solution or not.
This figures have been selected for demonstration because of fast convergence of solution, however, this is a relevant question.
Thus we do in-place decoding for the whole range of $\Delta T_i$ with appropriate witting time both in the encoding and the decoding process.
The map of encoded signals for both symmetric and asymmetric models have been shown already in figure~\ref{fig:t_t}.
The corresponding decoded signals are shown in figure~\ref{fig:decode}.
For both models there are some area that the signals are decodeable.

In the out-of-place decoding method, the signal can again be decoded for some periodic signals.
As we have a periodic signal, there is an additional degree of freedom, the phase.
That means when we are going to feed a periodic signal into an empty device, which droplet should go first.
We observe that for the periodic signals the result is sensitive to this phase (the index of the starting droplet).
Then in a periodic pattern, if we start decoding from different phases, we might get different signals.
The resulting signals are again periodic and the original signal may be among them.

\begin{figure}
\centering


\includegraphics[width=6cm]{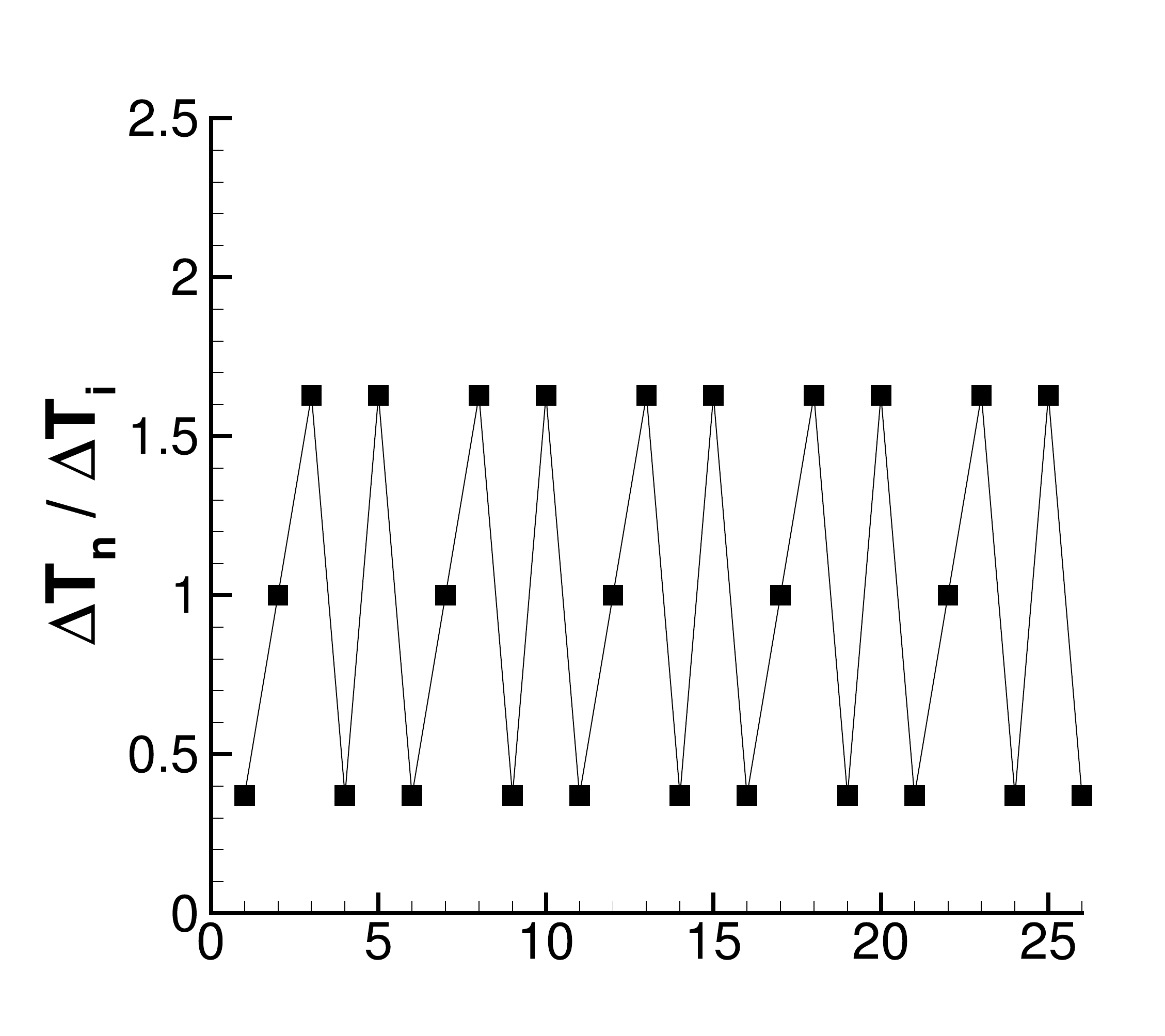} \\
\includegraphics[width=6cm]{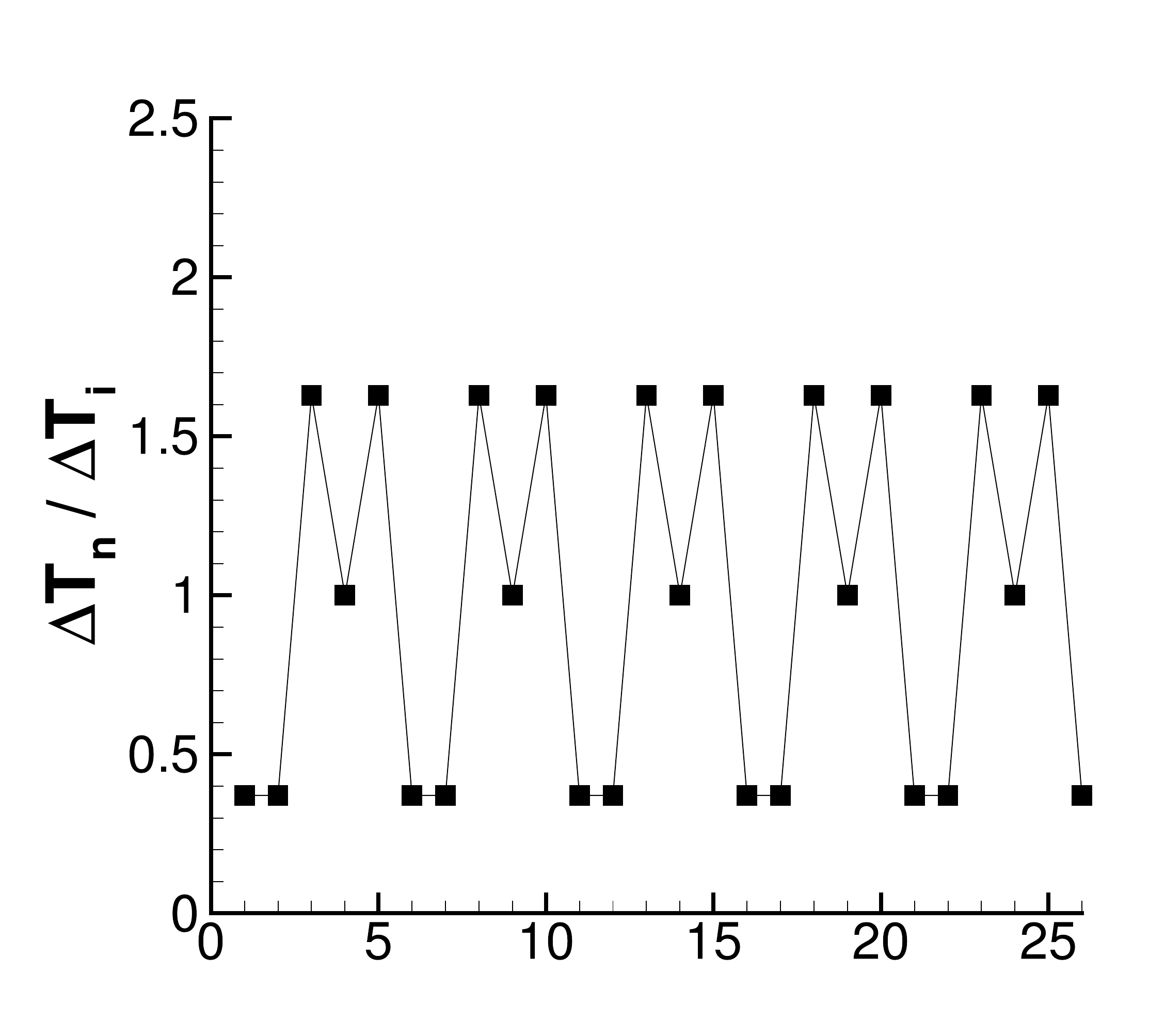} \\
\includegraphics[width=6cm]{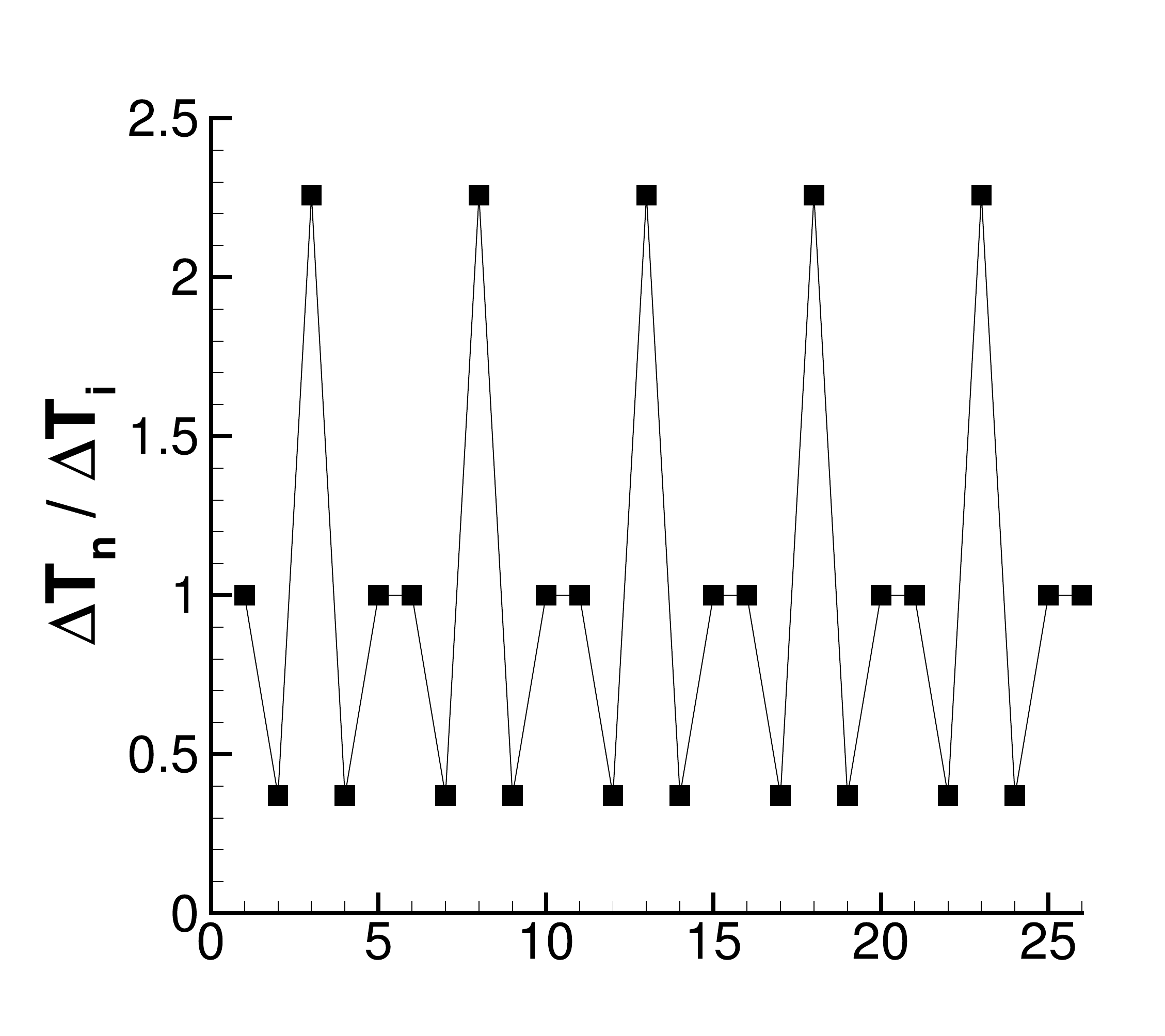} \\

\caption{\label{fig:revers_phase}
Out-of-place decoding: a periodic output signal (top) of asymmetric setup for $\Delta T_i= 1/15$ has a period of $5$.
The output depends on the phase of the signal in decoding process,  if we start from droplets $1$ (middle) or from  for $4$ (bottom).
If we start with the others (the numbers $2,3$ or $5$) the original uniform signal is reconstructed}
\end{figure}

An example is presented in figure~\ref{fig:revers_phase}.
The figure~~\ref{fig:revers_phase}a shows a periodic signal which is encoded by asymmetric setup with a flow of droplets with $\Delta T_i = 1/15$.
The numbers in the figure indicate the phase.
As the signal has a period of $5$, we can feed this sequence into an empty decoding device starting from any of this droplets.
As the figure shows different patterns are archived if we start decoding process with droplets numbered $1$ and $4$, while the original sequence is constructed if we start from the other three droplets.
The similar futures are observed for symmetric model, however, only the results of asymmetric are presented here.

Up to now we only studied the output signal.
We can go furthere and look at the intermediate pathway and study the patterns of pipe selections.
We can follow the train of the droplets in input we can follow them and see any droplet is choosing which branch,
say up ($\uparrow$) or down ($\downarrow$).
This give us another way of decoding the signal.
This definition of signal has been introduced by Jousse {\it et al.}~\cite{Bifurcation.of.droplet.PRE} in their experiment.
We shall try the same with our model and compare the results with the experiment.

Here we look at the signal in our asymmetric model and without lack of generality,
we suppose the shorter branch is labeled $\uparrow$.
If we count the number of droplets passing through ``up'' and ``down'' branches and show them by $n_\uparrow$ and $n_\downarrow$, respectively,
then $p =n_\uparrow /( n_\uparrow + n_\downarrow) $ and $q =n_\downarrow /( n_\uparrow + n_\downarrow) $ are probability of a bobble passing through ``up'' or ``down'' branches.
We expect in a long run, more droplets pass through the shorter branch ($\uparrow$) than longer ($\downarrow$),
then $1 \geq r = p - q \geq 0$.
But $r$ is not only related to the pipe lengths, but also it varies by $\Delta T_i$.
For a large enough $\Delta T_i$, when any droplet reaches an empty loop, all droplets pass through shorter branch and $r=1$.
Decreasing  $\Delta T_i$ in some point the droplets choose ``up'' and ``down'' branches alternatively.
Here we have $r =0$. One should note that, here the up-down signal is alternative and sequence of output intervals is still uniform and identical to input signal. Thus if up-down signal is periodic the sequence of time intervals is also periodic, but they don't need to have the same period.
As we have considered very small difference between the pipe lengths,
for any periodic up-down signal and in any period the difference between the numbers of $\uparrow$ and $\downarrow$ is $0$ or $1$.
That means for any periodic signal with period $P$, $r$ is either $0$ or $1/P$, for even and odd $P$ numbers, respectively.

\begin{figure}

\includegraphics[width=4cm]{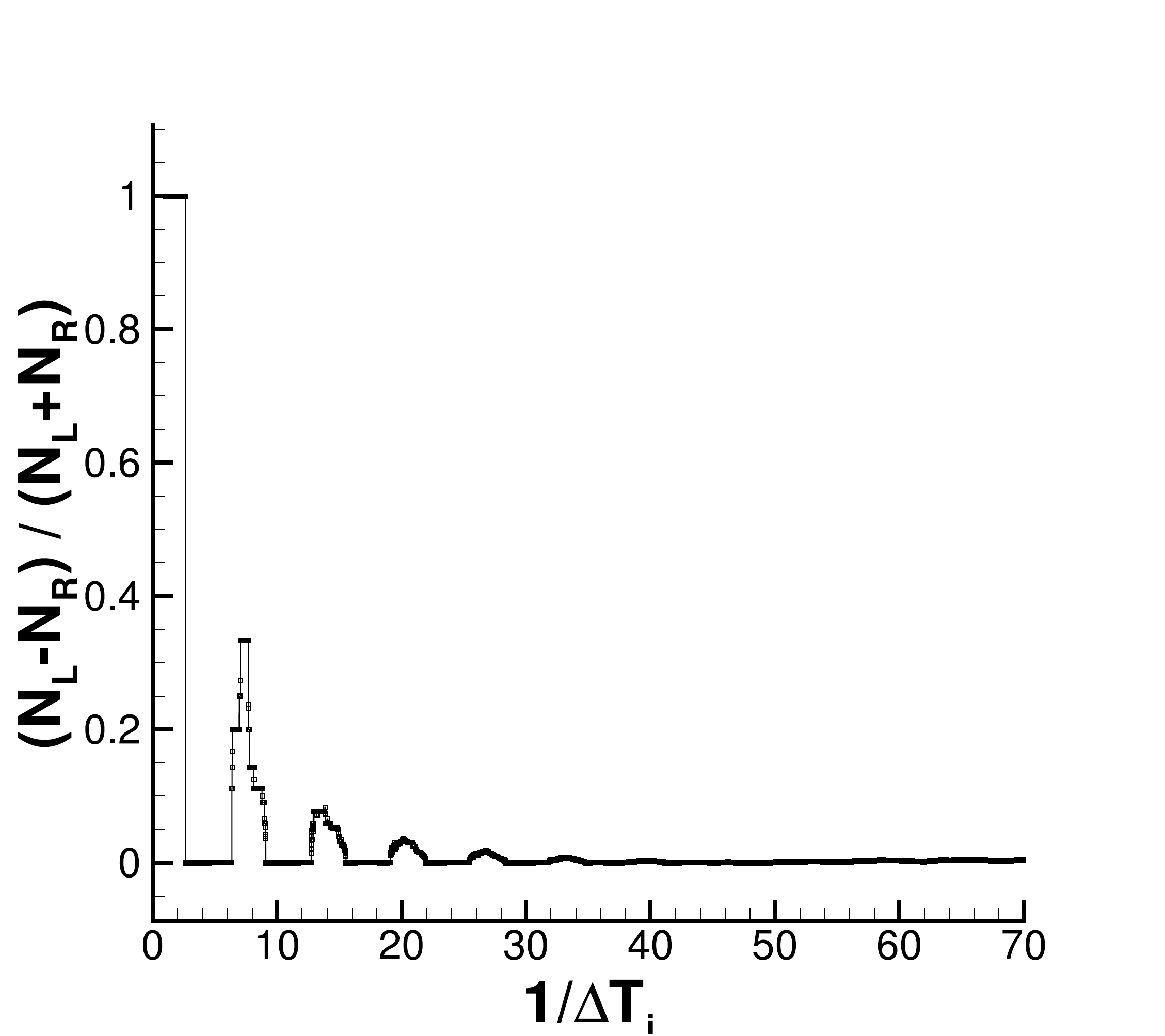}
\includegraphics[width=4cm]{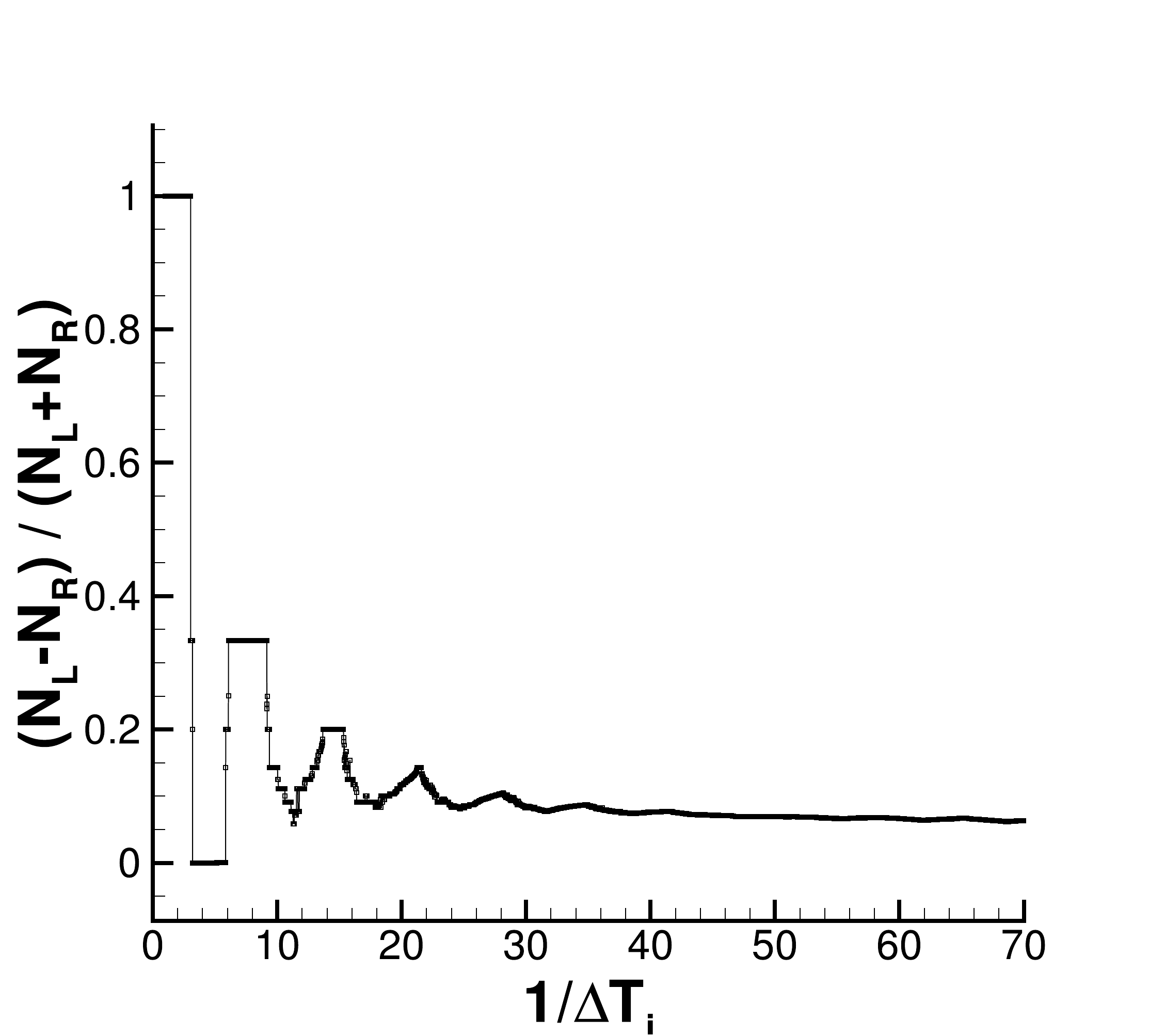}

\caption{\label{fig:n_ratio} Shows the ratio of the difference between the number of droplets passed through each pipe versus input time interval
$(n_0 - n_1)/(n_0 + n_1)$.
Symmetric setup (left), Asymmetric setup (right).
}
\end{figure}

\section{Conclusion}

In this work we used a simple setup of microfluidic network (FIG. \ref{fig:setup}) and a set of deterministic equations to simulate the passage of droplets or bubbles in the microfluidic network.
The fluid had a given constant flow. Which carries droplets. The time interval between successive droplets is constant at the input.
We found that patterns of droplets at the output depends on the time interval between the input droplets.
We observed that in general we have both periodic and chaotic patterns of output depending on the rate of entering droplets.

We found that although deterministic the system is not always reversible. We reversed the direction of time (or flow) in simulations and we found that in some situation the process is irreversible and droplets choose the other branch to come back which means a droplet leaving one of the branches, may choose the other one to return, if the time is reversed. This also depends on the input time interval of droplets.

We observed that there is a relation between irreversibility and chaos in our system. In general, when the output signal is periodic, it may be decoded to the original input signal by just reversing the flow direction. When it is chaotic, it is not possible to generate the input signal (decode) .

\bibliography{bubbles}

\begin{thebibliography}{14}
\expandafter\ifx\csname natexlab\endcsname\relax\def\natexlab#1{#1}\fi
\expandafter\ifx\csname bibnamefont\endcsname\relax
  \def\bibnamefont#1{#1}\fi
\expandafter\ifx\csname bibfnamefont\endcsname\relax
  \def\bibfnamefont#1{#1}\fi
\expandafter\ifx\csname citenamefont\endcsname\relax
  \def\citenamefont#1{#1}\fi
\expandafter\ifx\csname url\endcsname\relax
  \def\url#1{\texttt{#1}}\fi
\expandafter\ifx\csname urlprefix\endcsname\relax\def\urlprefix{URL }\fi
\providecommand{\bibinfo}[2]{#2}
\providecommand{\eprint}[2][]{\url{#2}}

\bibitem[{\citenamefont{Squires and Quake}(2005)}]{squires2005mfp}
\bibinfo{author}{\bibfnamefont{T.}~\bibnamefont{Squires}} \bibnamefont{and}
  \bibinfo{author}{\bibfnamefont{S.}~\bibnamefont{Quake}},
  \bibinfo{journal}{Reviews of Modern Physics} \textbf{\bibinfo{volume}{77}},
  \bibinfo{pages}{977} (\bibinfo{year}{2005}).

\bibitem[{\citenamefont{Whitesides}(2006)}]{whitesides2006oaf}
\bibinfo{author}{\bibfnamefont{G.}~\bibnamefont{Whitesides}},
  \bibinfo{journal}{NATURE-LONDON-} \textbf{\bibinfo{volume}{442}},
  \bibinfo{pages}{368} (\bibinfo{year}{2006}).

\bibitem[{\citenamefont{Choban et~al.}(2005)\citenamefont{Choban, Spendelow,
  Gancs, Wieckowski, and Kenis}}]{choban2005mlf}
\bibinfo{author}{\bibfnamefont{E.}~\bibnamefont{Choban}},
  \bibinfo{author}{\bibfnamefont{J.}~\bibnamefont{Spendelow}},
  \bibinfo{author}{\bibfnamefont{L.}~\bibnamefont{Gancs}},
  \bibinfo{author}{\bibfnamefont{A.}~\bibnamefont{Wieckowski}},
  \bibnamefont{and} \bibinfo{author}{\bibfnamefont{P.}~\bibnamefont{Kenis}},
  \bibinfo{journal}{Electrochimica Acta} \textbf{\bibinfo{volume}{50}},
  \bibinfo{pages}{5390} (\bibinfo{year}{2005}).

\bibitem[{\citenamefont{Schena et~al.}(1995)\citenamefont{Schena, Shalon,
  Davis, and Brown}}]{schena1995qmg}
\bibinfo{author}{\bibfnamefont{M.}~\bibnamefont{Schena}},
  \bibinfo{author}{\bibfnamefont{D.}~\bibnamefont{Shalon}},
  \bibinfo{author}{\bibfnamefont{R.}~\bibnamefont{Davis}}, \bibnamefont{and}
  \bibinfo{author}{\bibfnamefont{P.}~\bibnamefont{Brown}},
  \bibinfo{journal}{Science} \textbf{\bibinfo{volume}{270}},
  \bibinfo{pages}{467} (\bibinfo{year}{1995}).

\bibitem[{\citenamefont{Ko et~al.}(2003)\citenamefont{Ko, Yoon, Yang, Pyo,
  Chung, Kim, and Kim}}]{ko2003pbm}
\bibinfo{author}{\bibfnamefont{J.}~\bibnamefont{Ko}},
  \bibinfo{author}{\bibfnamefont{H.}~\bibnamefont{Yoon}},
  \bibinfo{author}{\bibfnamefont{H.}~\bibnamefont{Yang}},
  \bibinfo{author}{\bibfnamefont{H.}~\bibnamefont{Pyo}},
  \bibinfo{author}{\bibfnamefont{K.}~\bibnamefont{Chung}},
  \bibinfo{author}{\bibfnamefont{S.}~\bibnamefont{Kim}}, \bibnamefont{and}
  \bibinfo{author}{\bibfnamefont{Y.}~\bibnamefont{Kim}}, \bibinfo{journal}{Lab
  on a Chip} \textbf{\bibinfo{volume}{3}}, \bibinfo{pages}{106}
  (\bibinfo{year}{2003}).

\bibitem[{\citenamefont{Prakash and Gershenfeld}(2007)}]{Bubble.Logic.Scince}
\bibinfo{author}{\bibfnamefont{M.}~\bibnamefont{Prakash}} \bibnamefont{and}
  \bibinfo{author}{\bibfnamefont{N.}~\bibnamefont{Gershenfeld}},
  \bibinfo{journal}{Science} \textbf{\bibinfo{volume}{315}},
  \bibinfo{pages}{832} (\bibinfo{year}{2007}).

\bibitem[{\citenamefont{Fuerstman et~al.}(2007)\citenamefont{Fuerstman,
  Garstecki, and Whitesides}}]{Coding_Decoding.Science}
\bibinfo{author}{\bibfnamefont{M.~J.} \bibnamefont{Fuerstman}},
  \bibinfo{author}{\bibfnamefont{P.}~\bibnamefont{Garstecki}},
  \bibnamefont{and} \bibinfo{author}{\bibfnamefont{G.~M.}
  \bibnamefont{Whitesides}}, \bibinfo{journal}{Science}
  \textbf{\bibinfo{volume}{315}}, \bibinfo{pages}{828} (\bibinfo{year}{2007}).

\bibitem[{\citenamefont{Ajdari}(2004)}]{ajdari2004sfn}
\bibinfo{author}{\bibfnamefont{A.}~\bibnamefont{Ajdari}},
  \bibinfo{journal}{Comptes rendus-Physique} \textbf{\bibinfo{volume}{5}},
  \bibinfo{pages}{539} (\bibinfo{year}{2004}).

\bibitem[{\citenamefont{Marr and Munakata}(2007)}]{1284650}
\bibinfo{author}{\bibfnamefont{D.~W.} \bibnamefont{Marr}} \bibnamefont{and}
  \bibinfo{author}{\bibfnamefont{T.}~\bibnamefont{Munakata}},
  \bibinfo{journal}{Commun. ACM} \textbf{\bibinfo{volume}{50}},
  \bibinfo{pages}{64} (\bibinfo{year}{2007}), ISSN \bibinfo{issn}{0001-0782}.

\bibitem[{\citenamefont{Jousse et~al.}(2006)\citenamefont{Jousse, Farr, Link,
  Fuerstman, and Garstecki}}]{Bifurcation.of.droplet.PRE}
\bibinfo{author}{\bibfnamefont{F.}~\bibnamefont{Jousse}},
  \bibinfo{author}{\bibfnamefont{R.}~\bibnamefont{Farr}},
  \bibinfo{author}{\bibfnamefont{D.~R.} \bibnamefont{Link}},
  \bibinfo{author}{\bibfnamefont{M.~J.} \bibnamefont{Fuerstman}},
  \bibnamefont{and}
  \bibinfo{author}{\bibfnamefont{P.}~\bibnamefont{Garstecki}},
  \bibinfo{journal}{PHYSICAL REVIEW E} \textbf{\bibinfo{volume}{74}},
  \bibinfo{pages}{828} (\bibinfo{year}{2006}).

\bibitem[{\citenamefont{Wong et~al.}(1995{\natexlab{a}})\citenamefont{Wong,
  Radke, and Morris}}]{Equation1}
\bibinfo{author}{\bibfnamefont{H.}~\bibnamefont{Wong}},
  \bibinfo{author}{\bibfnamefont{C.~J.} \bibnamefont{Radke}}, \bibnamefont{and}
  \bibinfo{author}{\bibfnamefont{S.}~\bibnamefont{Morris}},
  \bibinfo{journal}{Journal of Fluid Mechanics} \textbf{\bibinfo{volume}{292}},
  \bibinfo{pages}{71} (\bibinfo{year}{1995}{\natexlab{a}}).

\bibitem[{\citenamefont{Wong et~al.}(1995{\natexlab{b}})\citenamefont{Wong,
  Radke, and Morris}}]{Equation2}
\bibinfo{author}{\bibfnamefont{H.}~\bibnamefont{Wong}},
  \bibinfo{author}{\bibfnamefont{C.~J.} \bibnamefont{Radke}}, \bibnamefont{and}
  \bibinfo{author}{\bibfnamefont{S.}~\bibnamefont{Morris}},
  \bibinfo{journal}{Journal of Fluid Mechanics} \textbf{\bibinfo{volume}{292}},
  \bibinfo{pages}{95} (\bibinfo{year}{1995}{\natexlab{b}}).

\bibitem[{\citenamefont{Bretherton}(1961)}]{Equation3}
\bibinfo{author}{\bibfnamefont{F.~P.} \bibnamefont{Bretherton}},
  \bibinfo{journal}{Journal of Fluid Mechanics} \textbf{\bibinfo{volume}{10}},
  \bibinfo{pages}{166} (\bibinfo{year}{1961}).

\bibitem[{\citenamefont{Gerami and Ejtehadi}(2000)}]{gerami2000hds}
\bibinfo{author}{\bibfnamefont{R.}~\bibnamefont{Gerami}} \bibnamefont{and}
  \bibinfo{author}{\bibfnamefont{M.}~\bibnamefont{Ejtehadi}},
  \bibinfo{journal}{The European Physical Journal B-Condensed Matter}
  \textbf{\bibinfo{volume}{13}}, \bibinfo{pages}{601} (\bibinfo{year}{2000}).

\end{thebibliography}

\end{document}